\documentclass{nature}
\usepackage{amsmath}
\usepackage[T1]{fontenc}
\usepackage[utf8]{inputenc}
\usepackage{graphicx}
\usepackage{amssymb}
\usepackage{dcolumn}
\usepackage{color}
\usepackage{bm}
\usepackage[normalem]{ulem}
\usepackage{url}
\usepackage{chemformula}

\usepackage{pdfpages}      


\begin{document}

\title{Quantized heat flow in moir\'e chern bands of bilayer graphene}

\author{Santanu Samai$^{1}$, Debangan Sarkar$^{1}$, Abhijit Halder$^{1}$, T. Taniguchi$^{2}$, K. Watanabe$^{2}$, Subroto Mukerjee$^{1}$, Saurabh Kumar Srivastav$^{3}$, Anindya Das$^{1}$\footnote{anindya@iisc.ac.in}}

\maketitle

\begin{affiliations}

\item Department of Physics, Indian Institute of Science, Bangalore, India.
\item National Institute of Material Science, Tsukuba, Japan
\item Department of Physics, Indian Institute of Technology (BHU), Varanasi, India

\end{affiliations}

\noindent
\textbf{When electrons are subjected simultaneously to a magnetic field and a periodic potential, they form the fractal Hofstadter spectrum, whose topological gaps host quantum Hall and Chern insulating states with distinct Chern numbers. While electrical transport has established the topology of these states, whether their heat transport is likewise universal has remained unexplored. Here, we measure the thermal conductance of quantum Hall, Chern insulator, and interaction-driven symmetry-broken Chern insulator states in a bilayer graphene-hexagonal boron nitride moir\'e superlattice with a moir\`e wavelength of $\sim$14 nm using Johnson-noise thermometry. We find that the thermal conductance ($G_Q$) is quantized in units of the thermal conductance quantum ($G_Q = t\kappa_0T$) and is determined solely by the Chern number ($t$), independent of the microscopic origin of the topological state. By directly revealing universal topological heat transport in Hofstadter bands, our work establishes thermal conductance as a stringent probe of moir\'e topological matter and provides a route to investigating more exotic phases, including fractional Chern insulators.}

\noindent\textbf{Introduction.}
\noindent The moir\'e superlattice structure with a long-range periodic potential, a period of few tens of nanometers, has emerged as a rich and versatile platform for exploring band topology\cite{song2019all,balents2020superconductivity,wu2017topological,andrei2020graphene}, strong correlations\cite{cao2018correlated,wu2018hubbard,li2021continuous,balents2020superconductivity} and superconductivity\cite{cao2018unconventional,lu2019superconductors,balents2020superconductivity} in two-dimensional materials like graphene-hexagonal boron nitride (hBN) superlattice~\cite{polshyn2020electrical,sun2021correlated}, twisted graphene\cite{cao2018unconventional,cao2018correlated}, and transition metal dichalcogenides (TMDCs)\cite{li2021continuous,xia2025superconductivity,guo2025superconductivity,wu2018hubbard}. Among these, graphene-hBN based superlattices\cite{ponomarenko2013cloning,wang2015evidence,hunt2013massive,spanton2018observation,kuiri2021enhanced} are textbook examples to study the motion of Bloch electrons in the presence of competing length scales of moir\'e periodic potential and magnetic length. The interplay between these two length scales drives the Bloch bands to form a recursive fractal energy spectrum known as Hofstadter's butterfly  \cite{hofstadter1976energy,ponomarenko2013cloning,dean2013hofstadter,hunt2013massive,wang2015evidence,spanton2018observation}. 
The Hofstadter spectrum gaps are parameterized by the normalized carrier density ($n/n_0$) and magnetic flux density ($\phi/\phi_0$), where $n$ and $\phi$ are the carrier density and magnetic flux, respectively, while $n_0$  corresponds to one electron per moir\'e unit cell and $\phi_0$ corresponds to magnetic flux quantum ($h/e$). The different topological orders of the fractal gaps are traced by the different straight lines in the normalized density-normalized flux plot, also known as the Wannier plots\cite{dean2013hofstadter,hunt2013massive,spanton2018observation}. Each straight line is modeled by the Diophantine equation $n/n_0=t\frac{\phi}{\phi_0}+s$\cite{dean2013hofstadter,hunt2013massive}. The set of numbers $(t,s)$ encodes the topology of the fractal gaps. The number $t$ is associated with the electrical Hall conductance, ${\sigma_{xy}}=t\frac{e^2}{h}$ (where, $e$ is the electronic charge and $h$ is Planck's constant) and determines the topology of the band. At single particle levels, in the graphene-hBN superlattice, the topological gaps with integers $t$ and $s=0$ are described as regular quantum Hall (QH) states. The topological gaps with $s=\pm4$ are also considered as regular QH states emerging from the moir\'e bands. The topological gaps with integers $t$ and $s$ ($0<s<\pm 4$) are classified as Chern insulators (CI). In addition to these single-particle topological gaps, strong electronic interactions can give rise to symmetry-broken Chern insulators (SBCI) with integer $t$ but fractional $s=p/q$ (where $p$ and $q$ are coprime integers), and further fractional Chern insulators (FCI) with both fractional $t$ and $s$.

\noindent Using electrical transport measurements, the experimental observation of a wide variety of topological phases emerging from the recursive Hofstadter fractal band structures have provided compelling evidence of QH, CI, SBCI and FCI through their quantized electrical conductance values\cite{hunt2013massive,wang2015evidence,das2021symmetry,saito2021hofstadter}. 
The CI states arise from a qualitatively different mechanism beyond the free-electron Landau levels in QH, which involves the moir\'e superlattice potential hybridizing the Landau
levels to form Hofstadter subbands with non-trivial Bloch wavefunctions and Berry curvature distributed across the moir\'e Brillouin zone. 
Further, for SBCI and FCI states, this reconstructed band structure is reorganized by electron-electron interactions, giving rise to topological order with no free-particle counterpart. 
Thermal transport has potential to reveal the details about the edge structures\cite{wen1992theory,wen1990chiral,Kane1994,kane1995impurity} and the nature of heat-carrying edge modes like possible edge reconstruction\cite{kane1996thermal,jezouin2013quantum,banerjee2017observed,banerjee2018observation,Srivastaveaaw5798,PhysRevLett.126.216803,srivastav2022determination,le2022heat,roy2026half}, or additional neutral excitations\cite{banerjee2018observation,PhysRevLett.126.216803,srivastav2022determination,kumar2022observation,kumar2024absence,kumar2024electrical}, which electrical measurements
are not sensitive to. Thus, thermal conductance measurements of CI, SBCI and FCI states are crucial to shed light on the details about the edge structures of Hofstadter fractal band, and test whether the bulk-edge correspondence for the heat transport survives beyond the conventional QH.  
Despite its great importance in elucidating the physics of the edge states of CI, SBCI, and FCI, thermal conductance in these systems remains largely unexplored experimentally.

\noindent Recently, thermal conductance ($G_Q$) measurements have been performed on a monolayer graphene-hBN superlattice\cite {zhang2026quantized}. However, rather than observing the quantized values, $G_Q = t\kappa_0T$ (where $\kappa_0=\pi^2k_\mathrm{B}^2/3h$, $k_\mathrm{B}$ is the Boltzmann constant, 
and $T$ is the temperature), expected from the number of topological edge modes $t$ (for integer $t$), the measured thermal conductance was reduced by one ballistic channel, and it was attributed to the heat Coulomb blockade (HCB) effect\cite{sivre2018heat}. As a result, the intrinsic quantized thermal conductance $G_Q = t\kappa_0T$ has yet to be experimentally demonstrated in moir\'e Chern bands. Here, we investigate the quantized thermal conductance of moir\'e Chern bands realized in bilayer graphene-hBN superlattices using highly sensitive Johnson-noise thermometry. We measure the thermal conductance of various integer quantum Hall states, Chern insulators, and interaction-driven symmetry-broken Chern insulators. Our results demonstrate that the thermal conductance is quantized ($G_Q = t\kappa_0T$), and determined solely by the corresponding topological invariant ($t$) across all the investigated topological states. These findings establish the universality of quantized heat transport in recursive Hofstadter fractal gaps, strengthen the connection between thermal response and topology, and provide new insights into the fundamental properties of topological quantum matter. 

\begin{figure*}
\centerline{\includegraphics[width=1\textwidth]{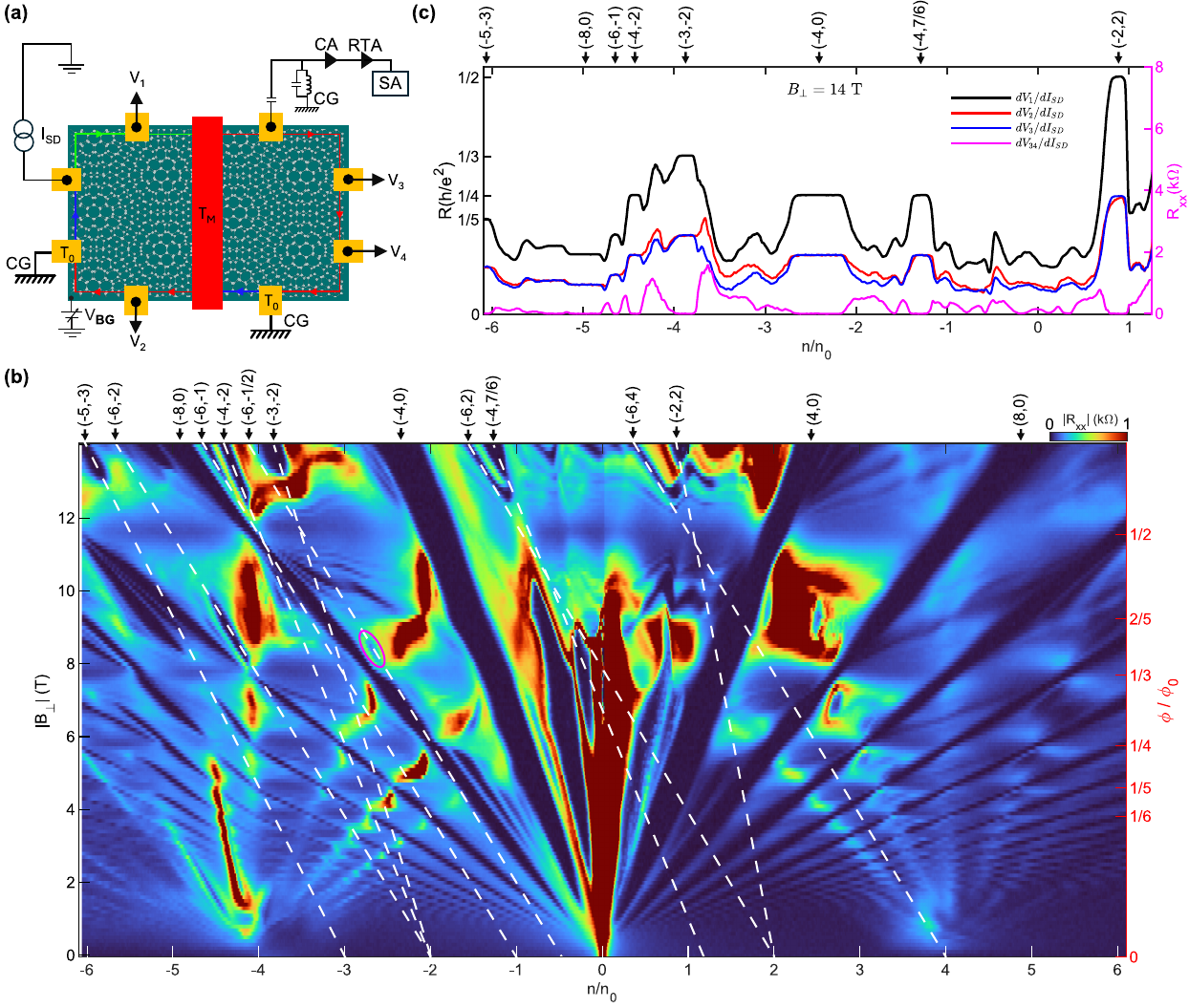}}
\caption{\textbf{Device schematic, measurement scheme, and QH to CI response of a hBN-BLG moir\'e superlattice.} (\textbf{a}) Schematic of measurement setup used for electrical transport and thermal conductance measurement. (\textbf{b}) 
$R_{xx} = dV_{34}/dI_{\mathrm{SD}}$ as a function of the carrier numbers per moir\'e unit cell, $n/n_0$, and perpendicular magnetic field, $B_{\perp}$ for Device~1. The right axis shows the magnetic flux per moir\'e unit cell, $\phi/\phi_0$. White dashed lines denote selected Diophantine trajectories corresponding to Chern insulating states, labelled by their Diophantine indices $(t,s)$. (\textbf{c}) Multi-terminal resistances measured in Device~1 at $B_{\perp}=14~\mathrm{T}$: $dV_1/dI_{\mathrm{SD}}$ (black), $dV_2/dI_{\mathrm{SD}}$ (red), $dV_3/dI_{\mathrm{SD}}$ (blue), and $dV_{34}/dI_{\mathrm{SD}}$ (magenta), plotted as a function of $n/n_0$. All these responses have been taken at base temperature, $T_{bath} \sim 20~mK$}

\label{Figure1}
\end{figure*}

\noindent\textbf{Device schematic and electrical transport for identifying QH, CI, and SBCI states:} 
The device schematic and measurement setup are shown in Figure~\ref{Figure1}(a). We fabricated an hBN-encapsulated, graphite-back-gated bilayer graphene (BLG) device, in which the BLG is precisely aligned with only one of the hBN layers. The top and bottom hBN layers have thicknesses of approximately $16-18$ nm. All electrical contacts, as well as the metallic floating reservoir (required for thermal conductance measurements\cite{jezouin2013quantum,banerjee2017observed,Srivastaveaaw5798,PhysRevLett.126.216803,srivastav2022determination}) located at the center of the device, were fabricated using one-dimensional edge contacts following hBN etching. As a result, the separation between the metallic reservoir and the graphite back gate is less than $7-8$ nm. For electrical transport measurements, we employ a standard low-frequency ($\sim31$ Hz) lock-in technique in constant-current mode with an excitation current of $2$ nA. We have repeated the measurement on two devices. Details of the fabrication and electrical measurements are described in Supplementary Note~1.

\noindent Figure~\ref{Figure1}(b) shows the 2D colormap of the quasi-longitudinal resistance, $dV_{34}/dI_{\mathrm{SD}}$, as a function of the carrier density per moir\'e unit cell, $n/n_0$, and perpendicular magnetic field, $B_{\perp}$, for Device 1. The right axis indicates the magnetic flux per moir\'e unit cell, $\phi/\phi_0$. Prominent dark-blue regions corresponding to incompressible states with vanishing $R_{xx}$ are clearly visible. With increasing magnetic field, a rich fractal structure emerges. In addition to the conventional QH states emanating from the charge neutrality point (CNP), numerous zero-$R_{xx}$ states appear away from the CNP. To identify these states, we overlay the Diophantine trajectories as white dashed lines for the prominent Chern insulating states, labelled by $(t,s)$, on which we performed thermal conductance measurements. The most prominent CIs include $(-2,2)$, $(-3,-2)$, $(-4,-2)$, $(-5,-3)$, $(-6,-1)$, $(-6,-2)$ and $(-6,4)$, among others. The corresponding 2D colormap of quasi $R_{xy}$ of Device~1 is shown in Supplementary Fig.~1(c), where the associated quantized Hall resistances can be clearly identified. The SBCI states $(-4,7/6)$ and $(-6,-1/2)$ in Device 1, along with $(-5,5/6)$ in Device 2 (see Supplementary Fig.~8), are also observed.

\noindent In Figure~\ref{Figure1}(c), we show line-cuts of four multi-terminal resistances measured in Device~1 at $B_{\perp}=14~\mathrm{T}$: $dV_1/dI_{\mathrm{SD}}$ (black), $dV_2/dI_{\mathrm{SD}}$ (red), $dV_3/dI_{\mathrm{SD}}$ (blue) and $dV_{34}/dI_{\mathrm{SD}}$ (magenta), plotted as a function of $n/n_0$. The black trace, $dV_1/dI_{\mathrm{SD}}$, can be treated as the Hall resistance, $R_{xy}$, in the absence of bulk conduction. The red and blue traces, $dV_2/dI_{\mathrm{SD}}$ and $dV_3/dI_{\mathrm{SD}}$, correspond to the reflected and transmitted components of the Hall resistance, denoted by $R_{xy}^{r}$ and $R_{xy}^{t}$, respectively. From these three responses ($R_{xy}$, $R_{xy}^{r}$, and $R_{xy}^{t}$), we find that $R_{xy}^{r}$ and $R_{xy}^{t}$ perfectly overlap at the QH, CI and SBCI plateaus, and each equals exactly half of $R_{xy}$, confirming the equipartition of the edge modes in our device, which is an important condition to be satisfied for thermal conductance measurements\cite{Srivastaveaaw5798,PhysRevLett.126.216803,srivastav2022determination}. The magenta trace, $dV_{34}/dI_{\mathrm{SD}}$, represents the longitudinal resistance, $R_{xx}$, in the absence of bulk conduction. The vanishingly small values of $R_{xx}$ at the QH, CI and SBCI states demonstrate the robustness of these states and the absence of any measurable bulk conduction, which is essential for extracting thermal conductance arising solely from the corresponding edge modes. The response for Device 2 remains also similar and shown in Supplementary Fig.~8.

\begin{figure*}
\centerline{\includegraphics[width=1\textwidth]{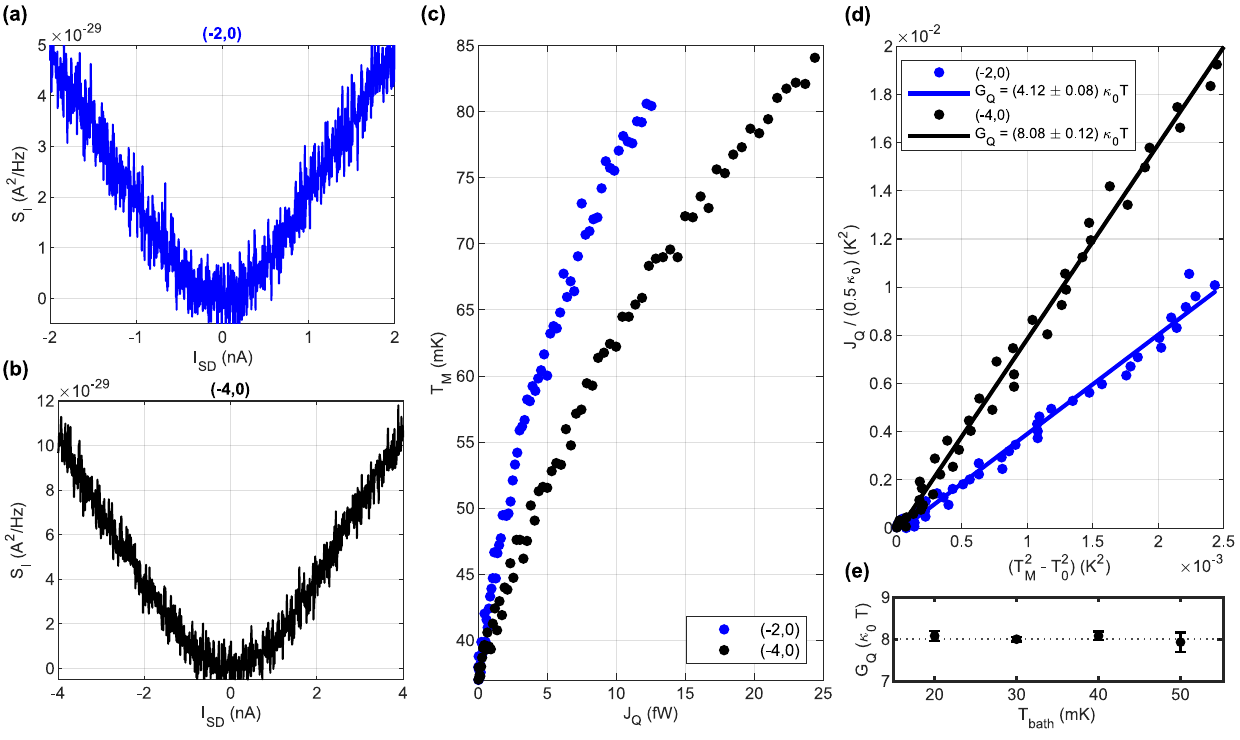}}
\caption{\textbf{Thermal conductance of conventional QH states in Device~1.}  \textbf{(a,b)} Excess thermal noise $S_{I}$ as a function of source current $I_{SD}$ for $(t,s)=(-2,0)$ (blue) \textbf{(a)}, $(t,s)=(-4,0)$ (black) \textbf{(b)}. (\textbf{c}) The temperature, $T_{M}$, of the floating-reservoir as a function of dissipated power, $J_{Q}$ for $(-2,0)$ $(N=4)$ (blue), $(-4,0)$ $(N=8)$ (black), where $N=2|t|$ denotes the total number of outgoing edge modes from the floating reservoir. (\textbf{d}) Scaled dissipated power, $J_{Q}/0.5\kappa_{0}$, plotted as a function of $T_{M}^2-T_{0}^2$ for the $(-2,0)$ and $(-4,0)$ at electron temperature $T_{0} \sim 37~mK$. Linear fits (solid lines) yield the values of thermal conductances, $G_{Q}=4.12\kappa_{0}T$  and $G_{Q}=8.08\kappa_{0}T$ for $(-2,0)$ and $(-4,0)$ states, respectively. (\textbf{e}) Measured thermal conductances, $G_{Q}$ of $(-4,0)$ state in Device~1 at different bath temperature, $T_{bath}$. Error bars corresponds to the fitting error.}
\label{Figure2}
\end{figure*}

\noindent \textbf{Thermal conductance measurement for QH states:} 
For the thermal conductance measurements, we follow our earlier works\cite{Srivastaveaaw5798,PhysRevLett.126.216803,srivastav2022determination,kumar2024absence,roy2026half}. Figure~\ref{Figure1}(a) illustrates the measurement scheme for a single chiral edge mode ($t = 1$). A dc source current, $I_{\mathrm{SD}}$, is injected from the source contact and propagates along the chiral edge channel (green line) into a central floating reservoir (FR), where it splits equally into two outgoing edge channels (red lines), which subsequently terminate at cold grounds maintained at the bath temperature, $T_0$. The power dissipated in the FR due to Joule heating is given by $J_Q=I_{\mathrm{SD}}^2/4tG_0$, where $t$ is the number of transmitting edge channels and $G_0=e^2/h$ is the quantum of electrical conductance. The electrons in the FR thermalize to a new steady-state temperature, $T_M$, which satisfies the heat balance relation\cite{jezouin2013quantum,banerjee2017observed,banerjee2018observation,Srivastaveaaw5798,PhysRevLett.126.216803,srivastav2022determination,kumar2024absence}: $J_Q=t\kappa_0(T_M^2-T_0^2)$. The electron temperature $T_M$ is determined from the excess thermal noise, $S_I=t k_B(T_M-T_0)G_0$\cite{Srivastaveaaw5798,PhysRevLett.126.216803,srivastav2022determination,kumar2024absence}, measured along the outgoing transmitted edge channel, as shown in Figure~\ref{Figure1}(a). To extract the thermal conductance, $G_Q$, we plot $J_Q$ as a function of $T_M^2-T_0^2$, where the slope directly yields $G_Q$.

\noindent Figure~\ref{Figure2}(a) and (b) show the measured excess thermal noise as a function of $I_{\mathrm{SD}}$ for the conventional QH states $(-2,0)$ and $(-4,0)$, respectively at $B_\perp = 7.5T$ and $T_0 \sim 37~mK$ in Device~1. The extracted electron temperature, $T_M$, as a function of $J_Q$, and $J_Q$ as a function of $T_M^2-T_0^2$ are plotted in Figure~\ref{Figure2}(c) and (d), respectively. The values of $G_Q$ extracted from the slopes (solid lines in Figure~\ref{Figure2}(d)) are $4.12\kappa_0T$ and $8.08\kappa_0T$ for the $(-2,0)$ and $(-4,0)$ states, respectively. The measured thermal conductance is expected to be $G_Q=2t\kappa_0T$ in our device geometry because the total number of outgoing edge modes from the FR is $2t$, consistent with our earlier works\cite{Srivastaveaaw5798,PhysRevLett.126.216803,srivastav2022determination}. Notably, we do not observe the effect of heat Coulomb blockade (HCB), which would reduce the measured $G_Q$ by one ballistic channel\cite{sivre2018heat}. This absence of HCB arises from the large capacitance between the FR and the graphite back gate due to the very thin hBN dielectric (see Supplementary Fig.~2). The detailed gain calibration and electron-temperature ($T_0$) determination are shown in Supplementary Note~2 and Supplementary Fig.~4,5,9, both of which are essential for accurately extracting $G_Q$. To further rule out HCB, we measure $G_Q$ for the $(-4,0)$ state in Device~1 as a function of bath temperature, and find that it remains nearly constant up to $50~\mathrm{mK}$, as shown in Figure~\ref{Figure2}(e).

\begin{figure*}
\centerline{\includegraphics[width=1\textwidth]{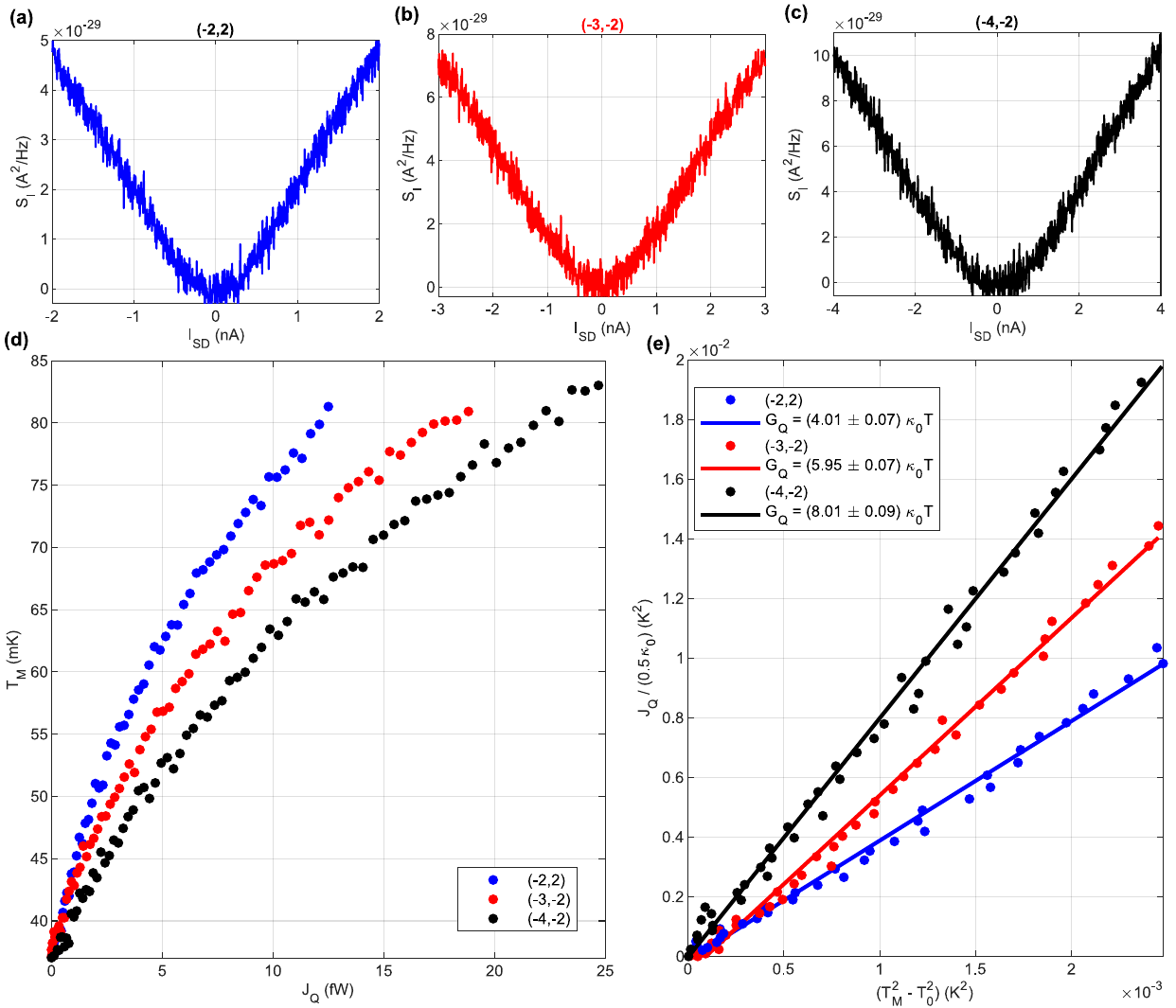}}
\caption{\textbf{Thermal conductance of Chern Insulator (CI) states in Device~1.}  \textbf{(a,b,c)} Excess thermal noise $S_{I}$ as a function of source current $I_{SD}$ for $(t,s)=(-2,2)$ (blue) \textbf{(a)},$(t,s)=(-3,-2)$ (red) \textbf{(b)}, $(t,s)=(-4,-2)$ (black) \textbf{(c)}. (\textbf{d}) $T_{M}$ as a function of $J_{Q}$ for $(-2,2)$ $(N=4)$ (blue), $(-3,-2)$ $(N=6)$ (red), $(-4,-2)$ $(N=8)$ (black), where $N=2|t|$ denotes the total number of outgoing edge modes from the floating reservoir. (\textbf{e}) Scaled dissipated power, $J_{Q}/0.5\kappa_{0}$, plotted as a function of $T_{M}^2-T_{0}^2$ 
at $T_{0} \sim 37~mK$. Linear fits (solid lines) yield values of thermal conductances of $G_{Q}=4.01\kappa_{0}T$, $G_{Q}=5.95\kappa_{0}T$ and $(G_{Q})=8.01\kappa_{0}T$ for the $(-2,2)$,$(-3,-2)$ and $(-4,-2)$ states, respectively.}
\label{Figure3}
\end{figure*}

\noindent \textbf{Thermal conductance measurement for CI states:} In Figure~\ref{Figure3}, we present the thermal conductance measurements for the CI states $(-2,2)$, $(-3,-2)$, and $(-4,-2)$ at $B_\perp = 14$ T and a base temperature of $T_0 \sim 37$ mK in Device~1. The corresponding robust conductance plateaus are shown in Figure~\ref{Figure1}(c). The measured excess thermal noise, $J_Q$ versus $T_M$, and the linear dependence of $J_Q$ on $T_M^2-T_0^2$ are shown in Figure~\ref{Figure3}(a--c), (d), and (e), respectively. The extracted thermal conductances are $G_Q = 4.01\kappa_{0}T$, $5.95\kappa_{0}T$, and $8.01\kappa_{0}T$ for the $(-2,2)$, $(-3,-2)$, and $(-4,-2)$ states, respectively, in excellent agreement with the expected quantized values, $G_Q = 2t\kappa_{0}T$, for the corresponding $(t,s)$ CI states. Similarly, the SBCI state $(-4,7/6)$ exhibits $G_Q = 8.04\kappa_{0}T$ (see Supplementary Fig.~6). Figure~\ref{Figure4} summarizes the measured $G_Q$ for several QH [$(\pm2,0), (-3,0), (\pm4,0)$] and CI [($\pm2,2), (-2,-2), (-2,-3), (-3,-2), (-4,-1), (-4,-2)$] states from two devices at different magnetic fields, all of which are consistent with their intrinsic quantized values of $G_Q = 2t\kappa_{0}T$.

\begin{figure*}
\centerline{\includegraphics[width=1\textwidth]{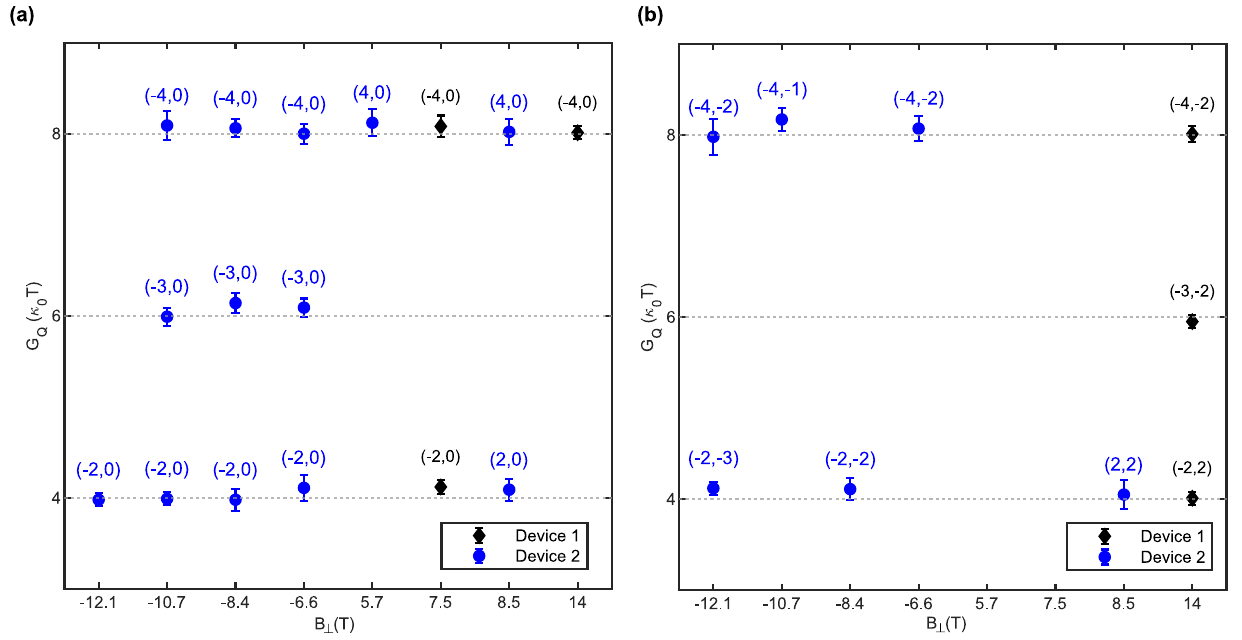}}
\caption{\textbf{Summary of the results measured in two devices.} (\textbf{a}) Measured $G_Q$ of several conventional QH states at different $B_{\perp}$. Black and blue solid circles represent data from Device~1 and Device~2, respectively, with error bars indicating the uncertainty in the extracted $G_Q$ values from fitting. 
(\textbf{b}) Measured $G_Q$, of CI states as a function of $B_{\perp}$. Black diamonds and blue solid circles correspond to Device~1 and Device~2, respectively. 
Error bars indicate the uncertainty in the extracted $G_Q$ values from fitting. All measurements were performed at the electron temperature, $T_0 \sim 37~\mathrm{mK}$ for Device 1 and $T_0 \sim 45~\mathrm{mK}$ for Device 2.
}
\label{Figure4}
\end{figure*}


\noindent\textbf{Discussion.} \ 
Both devices exhibit several CI states, although the thermal conductance measurements were limited to $(t,s)$ states with a maximum $|t|\simeq5$. For higher Chern numbers ($|t|=6$ and $8$), the measured thermal conductance remains below the expected quantized value, $G_Q=2t\kappa_{0}T$ (see Supplementary Fig.~13), most likely due to heat Coulomb blockade effect (see Supplementary Note~3). While HCB is expected to reduce the thermal conductance by maximum one ballistic channel, yielding $G_Q=(2t-1)\kappa_{0}T$\cite{sivre2018heat}, the measured value  for (-8,0) state is even slightly lower than this prediction. This may be arising due to the uncertainty in measuring $G_Q$ for states with a large number of edge modes ($N > 10$). In our devices, only a few symmetry-broken CI states, namely $(-4,7/6)$ and $(-5,5/6)$, were observed, and their thermal conductance was measured, which also remain consistent with $G_Q = 2t\kappa_{0}T$. The less number of SBCI states were observed likely due to the limited magnetic field of $B_\perp=14$ T. This is consistent with previous reports\cite{spanton2018observation}, where such SBCI states predominantly emerge at substantially higher magnetic fields.

\noindent\textbf{Conclusion.} \ 
In summary, we have experimentally demonstrated the universal quantization of thermal conductance in moir\'e Chern bands of hBN-bilayer graphene superlattice. Using sensitive Johnson-noise thermometry, we show that the heat carried by the edge states is determined solely by the Chern number, independent of whether the underlying topological state originates from conventional quantum Hall physics, single-particle Chern insulators, or interaction-driven symmetry-broken Chern insulators. Our observation of the intrinsic quantized thermal conductance, free from heat Coulomb blockade, establishes thermal transport as a universal probe of topology in Hofstadter bands. These results provide direct evidence for the bulk-edge correspondence in heat transport and open a pathway for exploring thermal signatures of more exotic topological phases, including fractional Chern insulators and non-Abelian states, in moir\'e quantum materials.

\newpage
\noindent \textbf{References}

\pagebreak
\section*{Materials and Methods}

\subsection{Device fabrication and measurement scheme:} Device fabrication began with the assembly of bottom-graphite-gated bilayer graphene (BLG) heterostructures encapsulated in hexagonal boron nitride (hBN), using conventional dry-transfer methods. Two distinct alignment schemes were used to generate moir\'e devices. For Device 1, the top and bottom hBN layers were sourced from a single crystal that had naturally cracked, leaving both resulting flakes with a shared crystallographic edge. During stacking, one straight edge of the top hBN was first matched to the longest edge of the BLG flake. The bottom hBN was then positioned such that its shared edge with the top hBN sat at an odd multiple of $30^\circ$. This approach deliberately aligns the BLG with only one hBN layer, preventing a double moir\'e pattern from forming. Device 2 used a different construction. The BLG flake was split into two segments: the first was aligned to the top hBN, and the second was rotated by $30^\circ$ relative to the first before being picked up. The bottom hBN was then added at roughly $15^\circ$ relative to the top layer, again to avoid generating a double moir\'e superlattice. Once stacking was complete, AFM measurements confirmed the thickness of the hBN layers. Contact regions were then patterned via electron-beam lithography (EBL). To expose graphene edges for 1D edge contacts, samples underwent reactive-ion etching (RIE) in a \ch{CHF3} and \ch{O2} gas mixture (40/4 sccm) at $25^\circ$C with 60 W RF power. This etch left both devices with a residual bottom-hBN thickness of about $\sim$7 nm. Metal contacts (Cr/Pd/Au, 6/12/65 nm) were then thermally evaporated under a base pressure near $1\times10^{-7}$ mbar, with excess metal removed via acetone/IPA lift-off. A second round of EBL and RIE defined the final device shape. Transport measurements were performed in a dilution refrigerator using standard low-frequency lock-in detection ($\sim$31 Hz) under constant-current excitation (2 nA). Thermal conductance was extracted via Johnson-noise thermometry, using an LC resonant circuit tuned to 730 kHz. The resulting noise signal passed through a custom cryogenic preamplifier at 4 K, then a room-temperature amplifier stage, before being read out on a spectrum analyzer. The full measurement schematic appears in Supplementary Fig. 3, with additional methodological details available in our earlier work~\cite{Srivastaveaaw5798,PhysRevLett.126.216803} and in the Supplementary Information.
 
\section*{Acknowledgements}
Authors thank Prof. Francois D. Parmentier, Prof. Yuval Gefen and Dr. Sourav Manna for useful discussions. A.D. thanks the Anusandhan National Research Foundation (ANRF) with project no: SP/ANRF-26-0244 and Department of Science and Technology (DST/NM/TUE/QM-5/2023) for the financial support. S.K.S. thanks Anusandhan National Research Foundation (ANRF) (ANRF/ARG/2025/009434/PS) for the financial support. K.W. and T.T. acknowledge support from the Elemental Strategy Initiative conducted by the MEXT, Japan and the CREST (JPMJCR15F3), JST. 

\section*{Author contributions}
S.S. contributed to device fabrication, measurement, data acquisition and analysis. D.S. and A.H. contributed in characterization of the devices. A.D. contributed in conceiving the idea and designing the experiment, data interpretation and analysis. K.W. and T.T. synthesized the hBN single crystals. S.K.S. and S.M. contributed to the discussion of the results and interpretation, and all the authors contributed to writing the manuscript.

\section*{Competing financial interests}
The authors declare no competing financial interests.

\section*{Data availability}
The data presented in the manuscript are available from the corresponding author upon request. 
 
\mbox{}
\includepdf[pages=-]{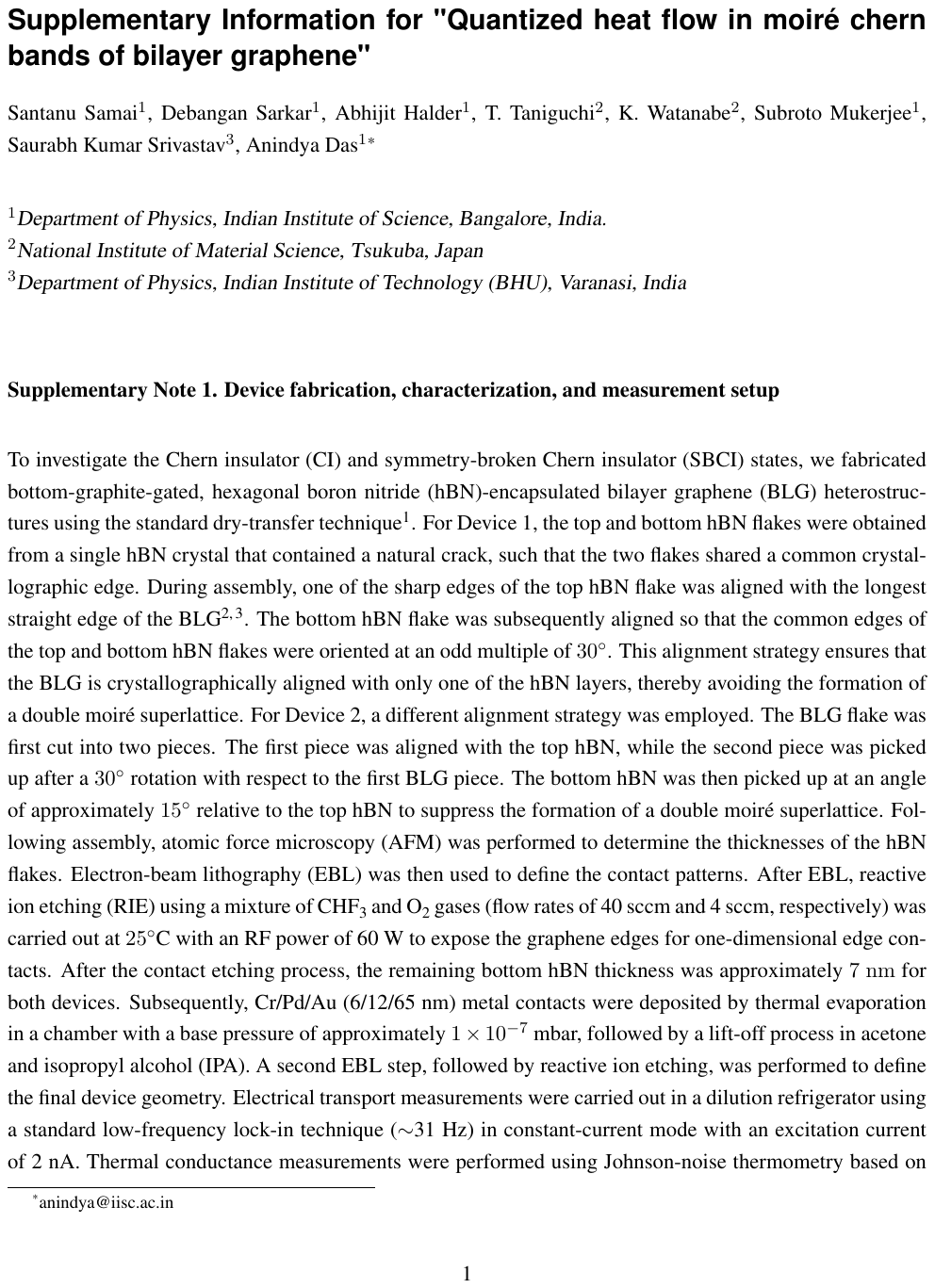}
 
\end{document}